\documentclass[aps,twocolumn,amsmath,amssymb,prl]{revtex4}
\usepackage{epsf}
\usepackage{graphicx}

\newcommand{\pphi}{\varphi}
\newcommand{\ppsi}{\Psi}
\newcommand{\vepsilon}{\varepsilon}
\newcommand{\mbf}[1]{\mbox{\boldmath $#1$}}
\renewcommand{\vec}[1]{{\mbf{#1}}}
\newcommand{\mvec}[1]{{\mbf{#1}}}

\newcommand*{\vsigma}{\mvec{\sigma}}

\begin{document}

\title{Triplet supercurrents in clean and disordered half-metallic ferromagnets} 

\author{Matthias Eschrig$^{1,\ast}$ and Tomas L\"ofwander$^{1,\dagger }$}

\affiliation{
$^{1}$Institut f{\"u}r Theoretische Festk{\"o}rperphysik and
DFG-Center for Functional Nanostructures,
Universit{\"a}t Karlsruhe, D-76128 Karlsruhe, Germany\\
$^{\ast}$e-mail: eschrig@tfp.uni-karlsruhe.de\\
$^{\dagger}$Present address: Department of Microtechnology and Nanoscience,
\mbox{Chalmers University of Technology, S-412 96 G\"oteborg, Sweden}
} 

\date{December 4, 2006; Published in Nature Physics {\bf 4}, 138-143 (2008).}

\begin{abstract}
{\bf 
Interfaces between materials with differently ordered phases present
unique opportunities to study fundamental problems in physics. One
example is the interface between a singlet superconductor and a
half-metallic ferromagnet, where Cooper pairing occurs between
electrons with opposite spin on one side, while the other displays
100\% spin polarisation. The recent surprising observation of a
supercurrent through half-metallic CrO$_2$ therefore requires a
mechanism for conversion between unpolarised and completely spin
polarised supercurrents.
Here we suggest a conversion mechanism based on electron spin
precession together with triplet pair rotation at interfaces with
broken spin-rotation symmetry. In the diffusive limit the triplet
supercurrent is dominated by inter-related odd-frequency $s$-wave and
even-frequency $p$-wave pairs. In the crossover to the ballistic
limit additional symmetry components become relevant.
The interface region exhibits a superconducting state of
mixed-spin pairs with highly unusual symmetry properties that opens
up new perspectives for exotic Josephson devices.
}
\end{abstract}

\maketitle

Half-metallic ferromagnets have great potential in the field of
spintronics as sources of spin-polarised electric currents.
Remarkably, they show conducting or insulating behaviour depending on
the direction of the electron spin.  Since only a few such half metals
are known, among them La$_{2/3}$Ca$_{1/3}$MnO$_3$ \cite{cha06} and
CrO$_2$ \cite{kei06}, their characterisation has attracted great
attention. Half metals, when contacted with other materials such as
superconductors, can also be used as well controlled test-laboratories
to study the interplay between different types of orders.

Recently, Keizer {\it et.al} \cite{kei06} reported a Josephson
supercurrent between two singlet superconducting electrodes (NbTiN)
separated by a wide region of CrO$_2$.  In a half metal only
conduction electrons with equal spin can be paired, since the other
spin species is insulating. Currently, the mechanism involved in the
conversion process between singlet Cooper pairs
$(\left|\uparrow\downarrow\right\rangle-
\left|\downarrow\uparrow\right\rangle) /\sqrt{2}$ and equal-spin pairs
$\left|\uparrow\uparrow\right\rangle$ at the interfaces between the
materials remains highly controversial. Moreover, the symmetries of
the relevant pairing correlations mediating the triplet supercurrent
and their dependence on the amount of disorder is 
currently debated.
It is claimed \cite{vol03,asa06} that the main source of the triplet
Josephson current in diffusive ferromagnets is odd-frequency $s$-wave
pairing amplitudes.  On the other hand, it has been shown that in
clean half metals also $p$-wave triplet pairing amplitudes are
important \cite{esc03}.

\begin{figure}[b]
  \includegraphics[width=\columnwidth]{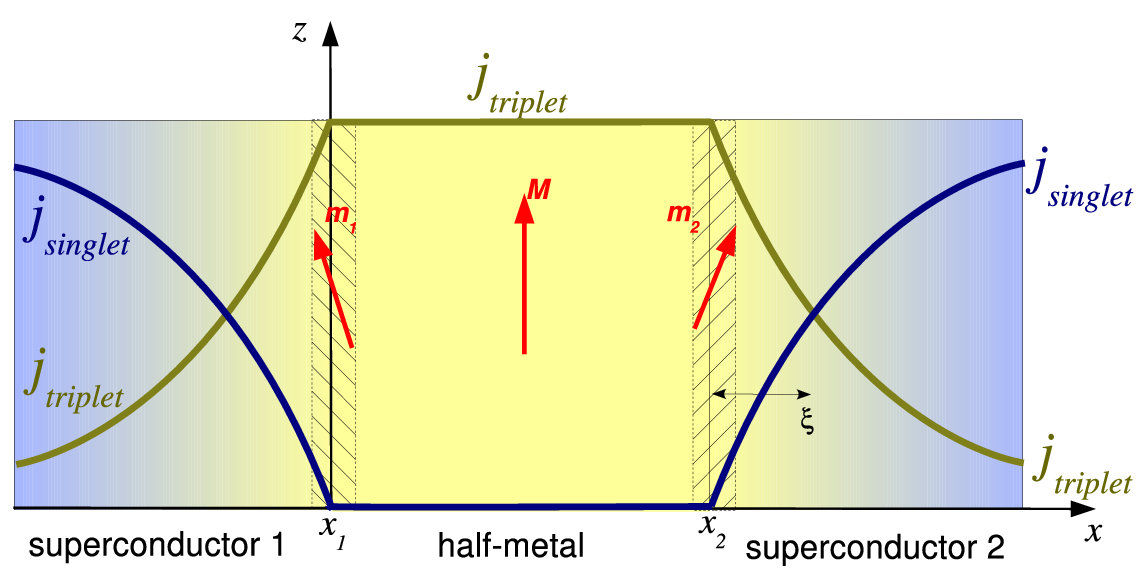}
  \caption{{\bf Conversion between singlet and triplet supercurrents.}
    We consider two singlet superconductor banks separated
    by a half-metallic ferromagnetic layer with magnetisation vector
    $\vec{M}$. Spin-rotation symmetry around $\vec{M}$ is broken at
    the interfaces (shaded), characterised by misaligned averaged interface
    moments $\vec{m}_1$ and $\vec{m}_2$.
    As a consequence, inside the superconductors 
    within a coherence length from the interfaces,
    there is a conversion from a supercurrent of singlet Cooper pairs
    ($j_{singlet}$, blue line) to a supercurrent of triplet Cooper pairs 
    ($j_{triplet}$, green line), as illustrated by the
        shading from blue to yellow. Only the triplet supercurrent
	can penetrate the half metal.
}
\label{fig:geometry}
\end{figure}

Consider the Josephson junction in Fig.~\ref{fig:geometry}.  It
consists of a half metal extending from $x_1$ to $x_2$, sandwiched
between two singlet superconductors. When a phase difference
$\chi_2-\chi_1$ exists between the superconducting order parameters,
an exotic form of Josephson effect occurs: a singlet supercurrent,
$j_{singlet}$ (blue in Fig.~\ref{fig:geometry}), is converted to an
equal-spin triplet supercurrent, $j_{triplet}$ (yellow in
Fig.~\ref{fig:geometry}), within an interface layer extending about a
superconducting coherence length into the electrodes. The equal-spin
triplet supercurrent flows through the half-metallic material, while
the singlet part is completely blocked. The sum of the singlet and
triplet currents is constant, obeying the continuity equation.

The mechanism of the current conversion we propose in this paper leads to a natural explanation of several
findings of the experiment \cite{kei06}: 
(i) a finite Josephson current in the half metal;
(ii) hysteretic shifts of the equilibrium phase
difference over the junction depending on the magnetic pre-history;
(iii) Josephson junctions involving half metals are $\pi$-junctions
after subtraction of the hysteretic shifts;
(iv) sample-to-sample fluctuations in
the magnitude of the critical current. 

\begin{table}[t]
  \includegraphics[width=0.8\columnwidth]{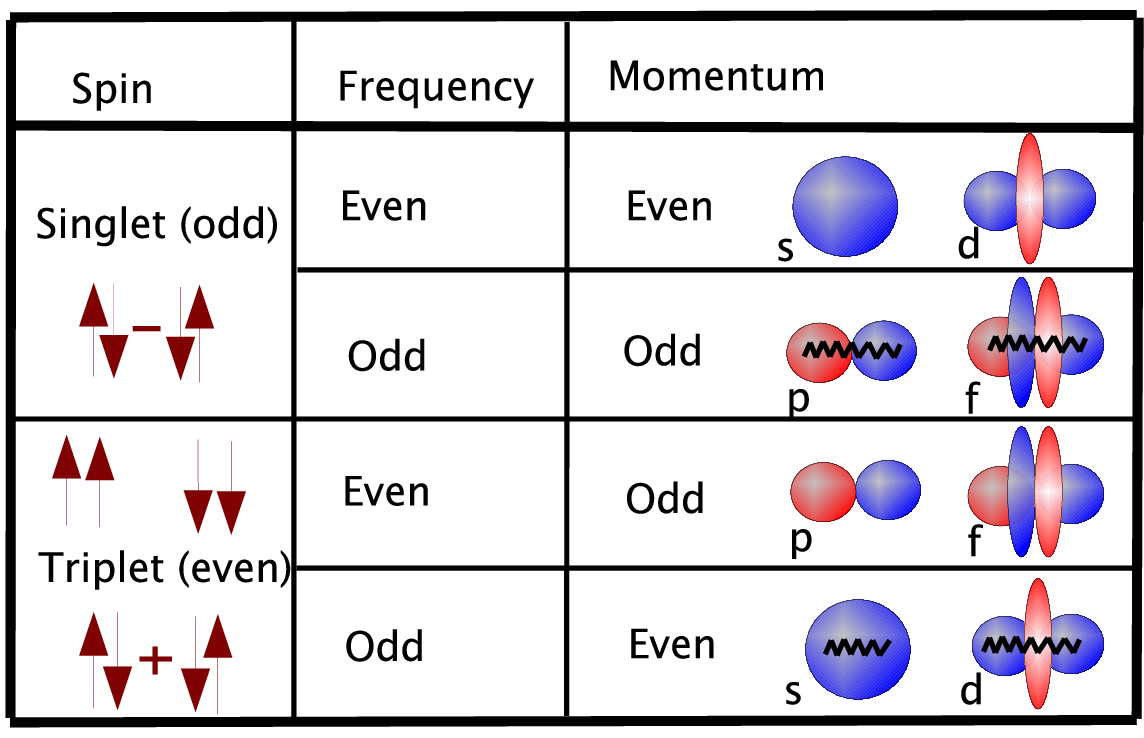}
  \caption{{\bf The four classes of superconducting correlations following
from the Pauli principle.}
All four symmetry components are induced in the superconducting regions 
next to the interface, but only the $\uparrow\uparrow$-triplet ones in 
the half metallic region. The dominating orbital contributions 
to the supercurrents 
in the half metal are shown in the lower two lines (triplet): 
even frequency $p$-wave and $f$-wave, and odd frequency $s$-wave and $d$-wave.
Wavy lines symbolise the dynamical nature of the odd frequency amplitudes.}
\label{fig:symmetries}
\end{table}

The four symmetry types of Cooper pairs allowed by the Pauli
exclusion principle \cite{tg07,fs06} are listed in
Table~\ref{fig:symmetries}.
We show that in moderately
disordered half metals the supercurrent is carried predominantly by
odd-frequency $s$-wave and $d$-wave amplitudes, multiplied with even
frequency $p$-wave and $f$-wave amplitudes. In the diffusive limit the
supercurrent is dominated by the product of the $s$-wave and the
$p$-wave amplitudes. We find that a peak in the temperature dependence
of the critical current, previously predicted for clean half metals
\cite{esc03}, is a robust feature also for disordered half metals.
We study the entire crossover from the ballistic to the diffusive regime.

\subsection{SPIN-MIXING AND TRIPLET-ROTATION}
The conversion process between the singlet and equal-spin triplet
supercurrents is triggered by two important phenomena taking place at
the interface: (i) spin mixing provides $S=1$, $m=0$ triplet
correlations near the interface, and (ii) breaking of spin-rotation
symmetry with respect to the magnetisation axis $\vec{M}$ in the half
metal allows for this $m=0$ triplet to be rotated into an $S=1$, $m=1$
triplet amplitude. 
Note that both phenomena are required for a non-vanishing Josephson effect. 

\begin{figure}[t]
  \includegraphics[width=0.6\columnwidth]{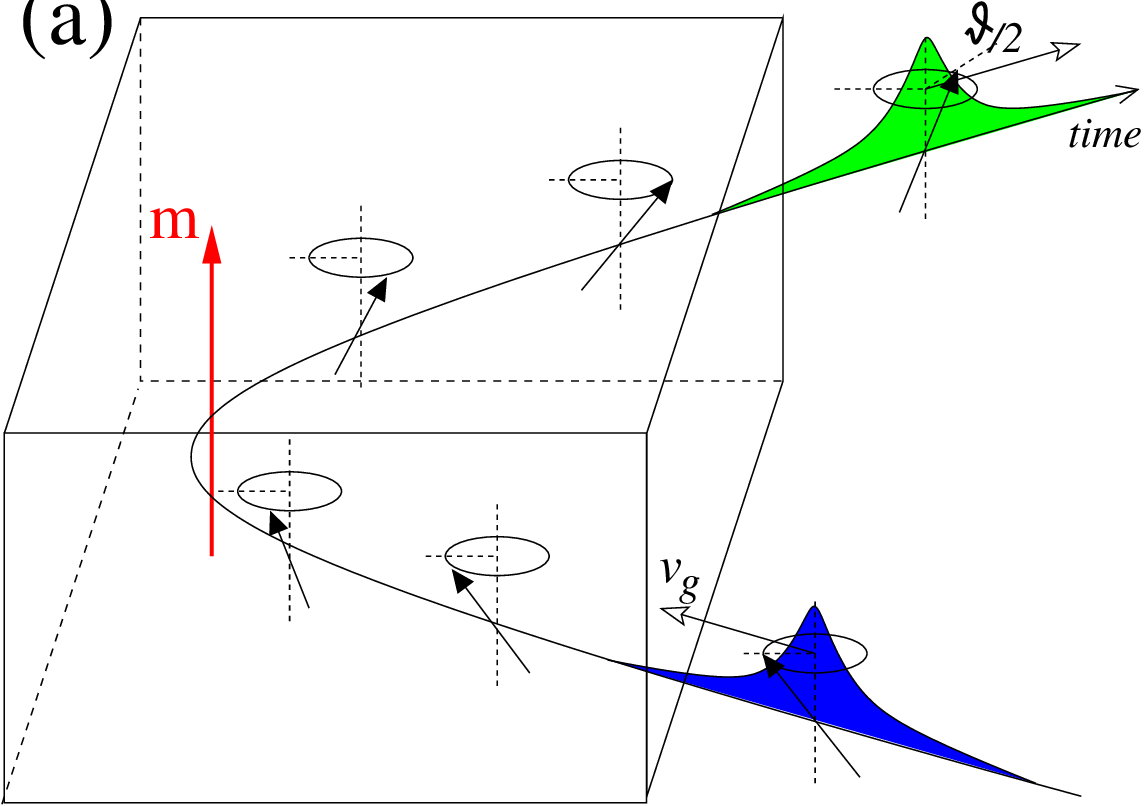}
  \includegraphics[width=0.35\columnwidth]{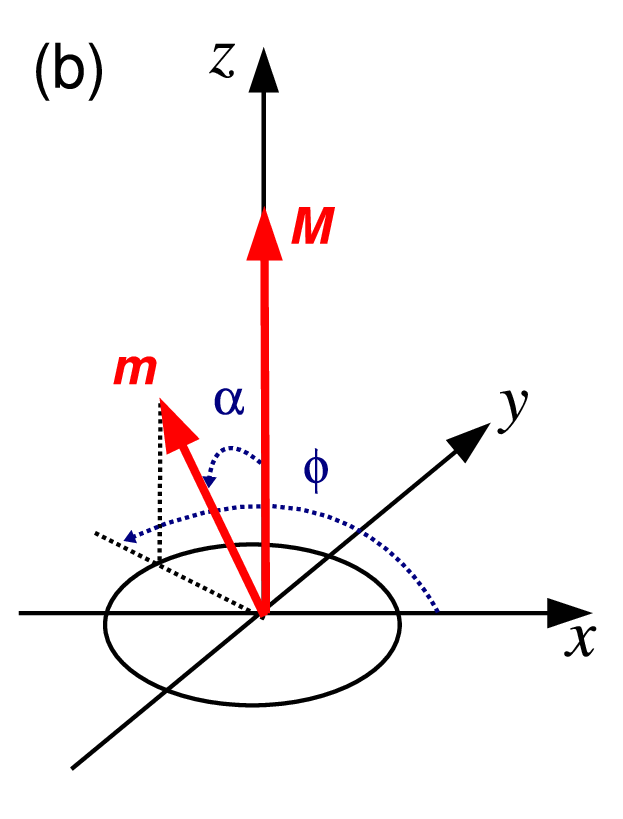}
  \caption{ {\bf Spin mixing and broken spin-rotation symmetry around
      $\vec{M}$.} (a) The spin-mixing angle $\vartheta $ corresponds
    to the precession of a spin around the magnetisation $\vec{m}$
    when a wave packet penetrates into the classically forbidden
    interface region. The spin component along $\vec{m}$ acquires a
    spin-dependent scattering phase.  (b) Definition of the polar
    angles $\alpha$ and $\varphi$ for an interface moment $\vec{m}$
    misaligned with respect to $\vec{M}$.}
\label{fig:spinmixing}
\end{figure}

Spin-mixing is the result of different scattering phase shifts that
electrons with opposite spin acquire when scattered (reflected or
transmitted) from an interface \cite{tok88,su91,esc04}. It results
from either a spin-polarisation of the interface potential, or
differences in the wave-vector mismatches for spin up and spin down
particles at either side of the interface, or both. It is a robust and
ubiquitous feature for interfaces involving strongly spin-polarised
ferromagnets.  Another, equivalent, way of discussing spin-mixing,
shown in Fig.~\ref{fig:spinmixing}(a), is in terms of a
spin-precession around $\vec{M}$ when wave packets penetrate the
interface region.

Broken spin-rotation symmetry leads to spin-flip processes at the
interfaces. Its origin is more subtle and deserves special attention.
We discuss in the Supplementary Information some possible origins of
misalignment of the interface moments with respect to the bulk
magnetic moment relevant for the material CrO$_2$. 
Magnetic materials often display surface order that is different from
the bulk order, possibly with disordered surface phases.
Here we only assume that the {\it averaged}
interface magnetic moment deviates from the direction of the bulk
magnetisation. A possible insulating magnetic interface layer is also
described by our theory. The exact microscopic distribution of local
moments at the interface is not important for superconducting
phenomena, since Cooper pairs are of the size of the coherence length
$\xi$ which is much larger than the atomic scale. It is, however,
important for the effective interface scattering matrix, as it can
lead to spin-flip terms if the distribution of the local-moment directions 
breaks spin-rotation symmetry around $\vec{M}$.

\begin{figure}[t]
  \includegraphics[width=1.0\columnwidth]{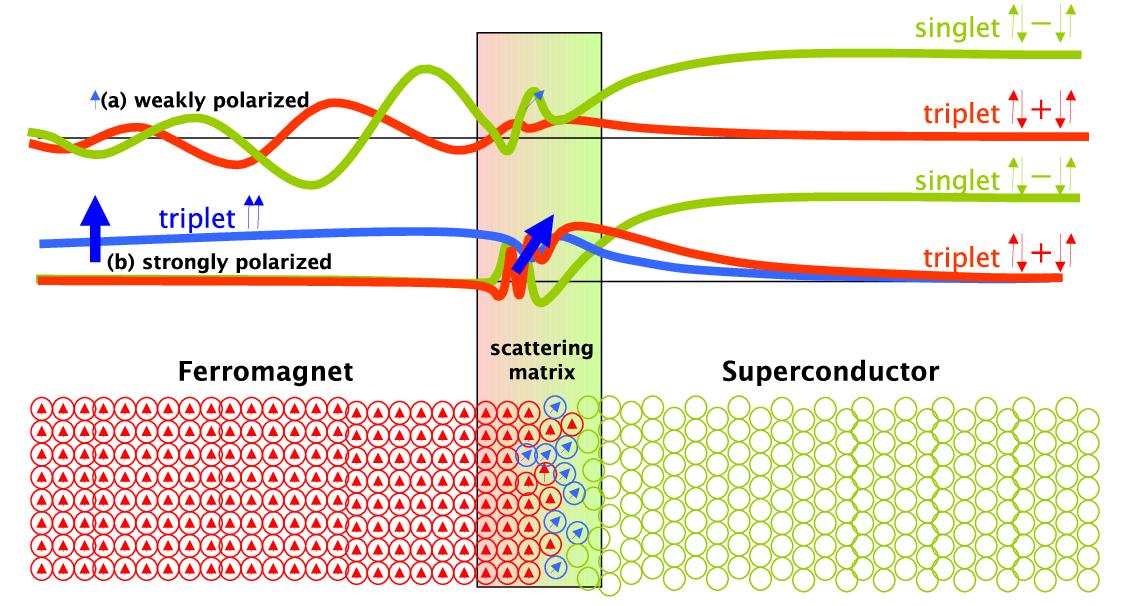}
  \caption{{\bf Proximity amplitudes induced at super\-con\-duc\-tor-ferromagnet
  interfaces.}
  (a) Misaligned spins in the interface region (described by a scattering matrix)
  add little to the main
  effect for weakly polarised ferromagnets, an out-of-phase
  oscillation of a singlet and a $\uparrow\downarrow+\downarrow\uparrow $ ($m=0$)
triplet component in the ferromagnet. 
  (b) For strongly polarised ferromagnets
  a considerable $m=0$ triplet amplitude is induced in the superconductor
  by the strong interface spin polarisation. Disordered interface moments
  rotate this $m=0$ into a $\uparrow \uparrow $ ($m=1$) triplet
  amplitude in the ferromagnet, if the averaged interface moment is misaligned
  with the bulk magnetisation.
  }
\label{fig:amplitudes}
\end{figure}

The two above-mentioned effects,
spin-mixing and broken spin-rotation symmetry around $\vec{M}$,
are interface properties
and lead to the appearance of the long-range $m=1$ triplet
correlations in the half metal as seen schematically 
in Fig.~\ref{fig:amplitudes}. This
is in contrast to the case of a weak ferromagnet coupled to a
superconductor, where without either large scale inhomogeneities
(e.g. domain wall structures near the interfaces) or strongly 
enhanced interface magnetism such correlations are negligible.

To quantify the above discussion, we employ a simple model that is
formulated in terms of an interface scattering matrix, which
connects incoming to outgoing waves in the asymptotic regions \cite{esc03}.
The scattering matrix depends in general on the following parameters:
(i) the total transmission $t$ between the
superconductor and the half metal;
(ii) the orientation of the 
averaged interface magnetic moment, $\vec{m}$, with polar angles
$\alpha $ and $\pphi $ as shown in 
Fig.~\ref{fig:spinmixing}(b);
(iii) two spin-mixing angles, one for reflection ($\vartheta $ )
and one for transmission ($\vartheta_{\uparrow\uparrow }$).
The most general form of the scattering matrix for the
tunnelling limit,
apart from irrelevant spin-independent phases, has the form 
(see the Supplementary Information)
\begin{equation}
\label{scatt}
\hat {\bf S}=
\left( \begin{array}{cc|c}
e^{\frac{i}{2}\vartheta} & 0& t_{\uparrow\uparrow}e^{i(\vartheta_{\uparrow\uparrow}+\frac{\vartheta }{4})}\\
0&e^{-\frac{i}{2}\vartheta} & t_{\downarrow\uparrow}e^{i(\vartheta_{\downarrow\uparrow}-\frac{\vartheta }{4})}\\
\hline
t_{\uparrow\uparrow}e^{-i(\vartheta_{\uparrow\uparrow}-\frac{\vartheta }{4})}&
t_{\downarrow\uparrow}e^{-i(\vartheta_{\downarrow\uparrow}+\frac{\vartheta }{4}) }
& - 1 \end{array} \right).
\end{equation}
Here, $t_{\uparrow \uparrow } = t \cos \frac{\alpha }{2}$
and $t_{\downarrow \uparrow } = t\sin \frac{\alpha }{2} $ are transmission
amplitudes from the two superconducting spin bands to the conducting
half-metallic spin-$\uparrow $  band, 
and $\vartheta $,
$\vartheta_{\uparrow \uparrow} $, 
and $\vartheta_{\downarrow \uparrow }= \pi+\pphi+\vartheta_{\uparrow \uparrow}$
are spin-mixing angles. 
Each interface $j=1$, $2$ is characterised by the five parameters 
$t_j$, $\alpha_j$, $\pphi_j$, $\vartheta_j$, and 
$\vartheta_{\uparrow \uparrow j}$, 
that in general can depend on the direction of
incoming quasiparticles. An alternative set is $t_{\uparrow \uparrow j}$,
$t_{\downarrow \uparrow j}$, $\vartheta_j$, $\vartheta_{\uparrow \uparrow j}$,
and $\vartheta_{\downarrow \uparrow j}$.
The presence of the spin-flip term
$t_{\downarrow \uparrow }e^{ i\vartheta_{\downarrow \uparrow }}$
in the scattering matrix, equation~(\ref{scatt}),
is a direct consequence of the broken spin-rotation symmetry 
around $\vec{M}$ at the interface.

\subsection{INDIRECT JOSEPHSON EFFECT}
In the following we calculate the Josephson current through the
junction to leading order in $t$ and $\vartheta$. This approximation
is not essential, but simplifies the following discussion while all
important phenomena are captured. The presence of an $m=0$ triplet
amplitude with a magnitude proportional to $\sin \vartheta$ [see
equation~(\ref{fsc}) below] is accompanied by a suppression of the
singlet pairing amplitudes proportional to $\sin^2 \! \frac{\vartheta }{2}$ in
the superconductors near the interface
(see Supplementary Table~S1), as illustrated in
Fig.~\ref{fig:amplitudes} (green lines) \cite{esc03,tok88,fs06}. It
leads to corrections to the singlet order parameter $\Delta$ that are
second order in $\vartheta$.
Thus, to leading order, the corresponding
suppression of $\Delta$ can be neglected. It follows that Anderson's
theorem \cite{And59,AG} holds and $\Delta$ is also insensitive to
impurity scattering (note, however, that in the {\it immediate }
interface region described by the scattering matrix the gap is
dramatically suppressed, e.g.  due to diffusion of magnetic moments;
this effect is included in our theory). For simplicity we consider the
case of equal gap magnitudes in the two superconductors,
$\Delta_j=|\Delta |e^{ i\chi_j}$, for superconductors $j=1$ and $j=2$,
see Fig.~\ref{fig:geometry}.

Due to spin mixing at the interfaces a spin triplet ($S=1$, $m=0$)
amplitude $f_{t_0j}(x)$ is developed that extends from the interfaces
about a coherence length into each superconductor,
\begin{eqnarray}
\label{fsc}
f_{t_0j}(x)=i\pi |\Delta | e^{i\chi_j} \sin \vartheta_j
\frac{ |\vepsilon_n| \ppsi_{0j}^s(x) + \Omega_n \ppsi_{0j}^a(x) }{\Omega_n^2},
\end{eqnarray}
where $\Omega_n=\sqrt{\vepsilon_n^2+|\Delta|^2}$. We have separated the
influence of the interfaces from that of the disorder in the
bulk materials by introducing the real functions $\ppsi_{0j}^{s,a}(x)$. The
superscript denotes symmetric ($s$) and antisymmetric ($a$) components
with respect to 
$\mu = \cos (\theta_p)$,
where $\theta_p$ is the angle between the Fermi velocity and
the $x$-axis. 
In the clean limit, 
$\ppsi_{0j}^a(x)=-\frac{\mbox{\small sgn}(\mu )}{2} e^{-|x-x_j|/\xi_S |\mu |} $ and
$\ppsi_{0j}^s(x)=\frac{\mbox{\small sgn}(\vepsilon_n )}{2} e^{-|x-x_j|/\xi_S |\mu |} $
where $\xi_S = v_{S}/2\Omega_n$, and $v_{S} $ is the Fermi
velocity in the superconductor.  For an arbitrary impurity concentration
the $\ppsi$-functions are modified and must be calculated
numerically for each given value of mean free path 
(see Supplementary Fig.~S1). 

The induced $m=0$ triplet amplitude derived above,
together with the presence of
spin-flip tunnelling amplitudes, leads to an equal-spin ($m=1$) pairing
amplitude $f_{\uparrow\uparrow}(x)$ in the half metal. 
The singlet component in the
superconductor, being invariant under rotations around any
quantisation axis, is not directly involved in the creation of the
triplet in the half metal. A picture of an indirect Josephson effect
emerges, therefore, that is mediated by the appearance of the $m=0$
triplet amplitudes in the superconductor.

In the tunnelling limit it is convenient to split the pairing amplitude
in the half metal into contributions induced at the left and right
interfaces:
$f_{\uparrow\uparrow}=f_{\uparrow\uparrow 1}+f_{\uparrow\uparrow 2}$,
with momentum-symmetric and antisymmetric components
\begin{equation}
\label{fhm}
f^{s,a}_{\uparrow\uparrow j}(x)= 2 \pi i 
A_j
|\Delta| e^{i\bar \chi_j }
\; \frac{|\vepsilon_n| 
}{\Omega_n^2}
\; \ppsi^{s,a}_{j}(x)
,
\end{equation}
where the amplitude is given by 
\begin{equation}
\label{amp}
A_j=2 t_{\uparrow\uparrow j} t_{\downarrow\uparrow j} \sin \left(\frac{\vartheta_j}{2}\right)
= t_j^2 \; \sin (\alpha_j ) \; \sin \left(\frac{\vartheta_j}{2}\right)
,
\end{equation}
and the effective phase by
\begin{equation}
\label{Jphase}
\bar \chi_j  = 
\chi_j +(\vartheta_{\uparrow\uparrow j}-\vartheta_{\downarrow\uparrow j})
= \chi_j - (\pi+ \pphi_j)
.
\end{equation}
In equation~(\ref{fhm}), we have separated the contributions from the
interface scattering and the contributions from the disorder in the
half metal by introducing the (real) functions $\ppsi^{s,a}_{j}$.

The Josephson current reads [see also equation~(S13) in the Supplementary Information]
\begin{equation}
\label{J1}
J_x=-J_c \sin (\bar \chi_2-\bar \chi_1),
\end{equation}
where the critical current density is given by
\begin{equation}
\label{Jc}
J_c= 
J_0
\frac{T}{T_c}\sum_{\vepsilon_n>0} 
\frac{|\Delta |^2\vepsilon_n^2}{\Omega_n^4}
\Big\langle \mu A_1A_2 (\ppsi^s_{2} \ppsi^a_{1} - \ppsi^s_{1} \ppsi^a_{2} )\Big\rangle .
\end{equation}
Here, the current unit is
$J_0=4\pi ev_{H}N_{H } T_c$, 
$N_{H }$ is the density of states at the Fermi level 
in the half metal, $e$ is the electron
charge, 
and $\langle \cdots \rangle= \int_0^1 d\mu \cdots $.

Equations~(\ref{amp})-(\ref{Jc}) describe an exotic Josephson effect in
several respects. Equation~(\ref{Jphase}) is related to the phase
dependence of the Josephson effect and can be tested for example by
studying the magnetic field dependence of the critical current. For a
half metal, there can be extra phases that lead to shifts of the usual
Fraunhofer pattern \cite{fs06,naz06}.  Within our model there are
contributions $\delta \pphi=\pphi_2-\pphi_1$ to the phases that depend on the
microscopic structure of the disordered magnetic moments at the two
interfaces. In particular, if the averaged magnetic interface moments
$\vec{m}_1$ and $\vec{m}_2$ are non-collinear in the plane
perpendicular to $\vec{M}$, such phases arise. The microstructure can
be affected for example by applying a magnetic field that leads to
hysteretic shifts $\delta \pphi (H)$ of the equilibrium positions 
depending on the
magnetic pre-history. When subtracting the shifts, the junction shows
the typical characteristics of a $\pi$-junction \cite{Buz82}, as
revealed by the minus-sign in equation~(\ref{J1}).  The possibility to
manipulate the shifts $\delta \pphi $ with an external field 
yields a way to measure
the relative orientation of $\vec{m}_1$ and $\vec{m}_2$ at the two
interfaces. Finally, the critical Josephson current is proportional to
the sine of the spin-mixing angles $\vartheta_j/2$, the transmission
probabilities $t_j^2$, and the sine of the angles $\alpha_j$ between
$\vec{m}_j$ and $\vec{M}$.  
This points to a strong sensitivity of the
critical Josephson current to interface properties and is expected to
lead to strong sample-to-sample variations. 
Note that none of the above parameters need to be small, such that 
critical currents of the order of that for normal junctions are
possible. All these findings are in
agreement with the experiment \cite{kei06}.

\subsection{ROLE OF DISORDER}
\begin{figure}
  \includegraphics[width=\columnwidth]{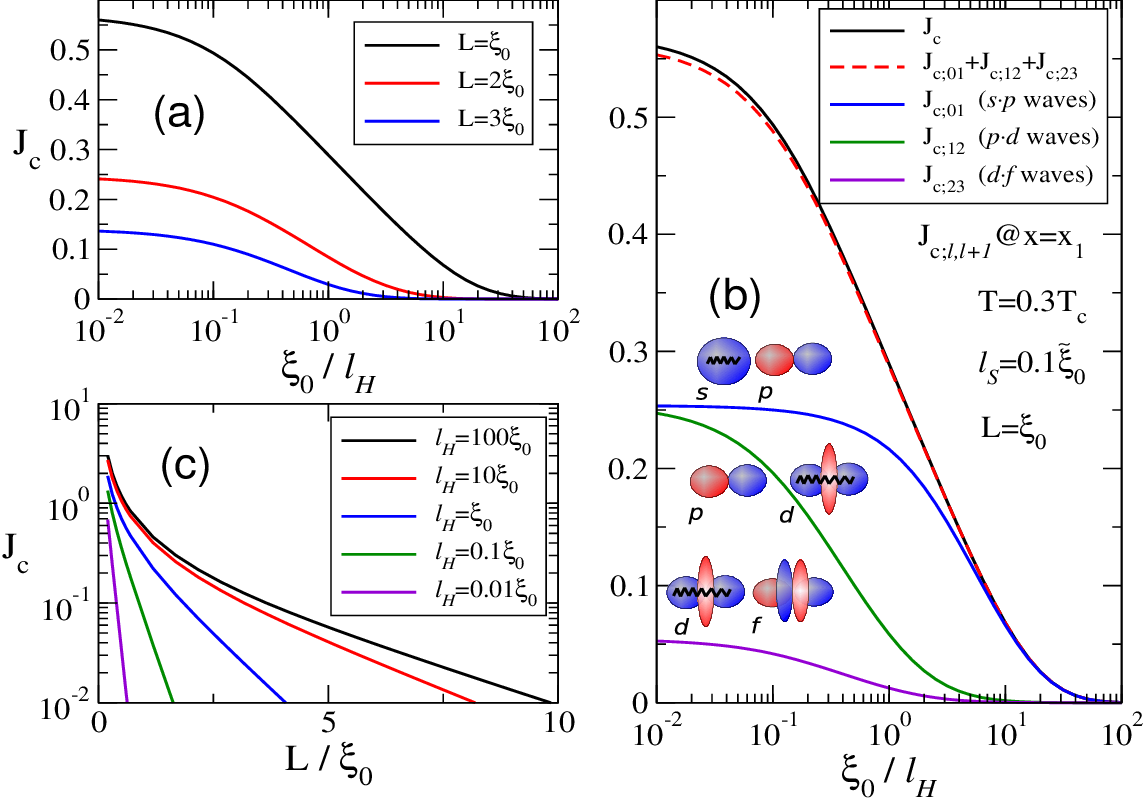}
  \caption{ {\bf Critical Josephson current $J_c$.} 
  (a) 
  In the crossover from the ballistic to the diffusive limits 
  $J_c$ is monotonically suppressed. 
  (b) 
  In the crossover region, contributions to $J_c$ from higher 
  partial waves ($l\geq 2$, $d$-wave, $f$-wave, etc.) 
  are suppressed. 
  In the diffusive limit the current (black line) is given by a product of
  $s$-wave ($l=0$) and $p$-wave ($l=1$) components (blue line). 
  (c)
  For given mean free path $\ell_H$, $J_c$ is exponentially suppressed 
  with junction length $L$.
  The unit for $J_c$ is $J_0 A_1A_2/4\pi $.
  The length unit is $\xi_0=v_{H}/2\pi T_c$. 
  We assumed an anisotropy of $t$ proportional to $|\mu |$.
  }
  \label{fig:mfp}
\end{figure}

We now proceed with a detailed description of the role of disorder in
the materials. 
For definiteness, in the remaining discussion we 
keep the mean free path in the superconducting banks fixed to
$\ell_S=0.1 \tilde \xi_0$ with $\tilde\xi_0= v_S/2\pi T_c$,
and vary the mean free path of the half metal.
It is well known that anisotropic superconducting correlations are sensitive to impurity
scattering. Studies of unconventional superconductivity reveal
that superconductivity disappears at a critical impurity
concentration \cite{Lar65,su91}. This is however not the case for the
proximity induced pairing amplitudes studied here. In
Fig.~\ref{fig:mfp} we show results for the critical Josephson current
as function of the elastic mean free path, normalised to
$\xi_0=v_{H}/2\pi T_c$. As shown in Fig.~\ref{fig:mfp}(a), the
critical current is monotonously suppressed for decreasing mean free
path, from the ballistic (left part of the abscissa in the figure) to
the diffusive (right part) limits. The suppression is exponential in
the diffusive limit, with a crossover taking place at a mean free
path $\ell_H$ comparable with the clean limit coherence length
$\xi_c=v_{H}/2\pi T$.

The critical Josephson current can be rewritten as a sum of terms, 
each consisting of products of neighbouring momentum symmetry 
components of the functions
$\overline \ppsi_j \equiv A_j\ppsi_j$ in equation~(\ref{Jc}), i.e.
$J_c=\sum_{l=0}^\infty J_{c;l,l+1}$, where $l=0,1,2,3,\ldots $ denotes
the \mbox{$s$-,} \mbox{$p$-,} \mbox{$d$-,} $f$-wave etc. pairing
components (see the Supplementary Information).
We have verified (see 
Supplementary Fig.~S2) that for ballistic systems the $p$-wave
amplitudes are larger than the $s$-wave amplitudes near the
interfaces, while the opposite holds for diffusive systems. The
amplitudes are tied to each other through the following general
relation between the momentum-antisymmetric and the momentum symmetric
parts:
$f_{\uparrow\uparrow}^{a}= -\mbox{sgn} (\vepsilon_n) \mu \xi_H 
\partial_x f_{\uparrow\uparrow}^{s} $,
where $\xi_H^{-1}= \ell_H^{-1}+ 2|\vepsilon_n |/v_{H} $. In the
diffusive limit, there is an additional relation,
$f_{\uparrow\uparrow}^{p{\rm -wave}}= -\mbox{sgn} (\vepsilon_n) \ell_H
\partial_x f_{\uparrow\uparrow}^{s{\rm -wave}} $.
It follows that the {\it magnitudes} of the amplitudes differ (their
ratio depends on the amount of disorder) while the {\it decay lengths}
of the two are always identical, crossing over from the ballistic
coherence length $\xi_c=v_{H}/2\pi T$ to the diffusive coherence
length $\xi_d=\sqrt{\ell_H \xi_c/3}$.  

The first three terms of the partial wave expansion of the
critical current are shown in Fig.~\ref{fig:mfp}(b).
The sum of these
contributions (red dashed line), composed of the $s\cdot p $
(blue), $p \cdot d$ (green), and $d\cdot f$ (purple) components,
amounts already to almost the entire current (black line). In the
diffusive limit, the current is carried almost exclusively by the
product of the even-frequency $p$-wave and the odd-frequency $s$-wave
pairing amplitudes (blue). In the crossover region to ballistic transport
there is an onset of contributions from higher order partial waves
$l\geq 2$. It is clear from the figure, that for $\ell_H \ge \xi_0$
the diffusive Usadel approximation breaks down.
Note that in the half metal only partial waves compatible with the spin triplet 
combinations of Table~\ref{fig:symmetries} are possible, as indicated
in Fig.~\ref{fig:mfp}(b).

\begin{figure}[t]
\includegraphics[width=\columnwidth]{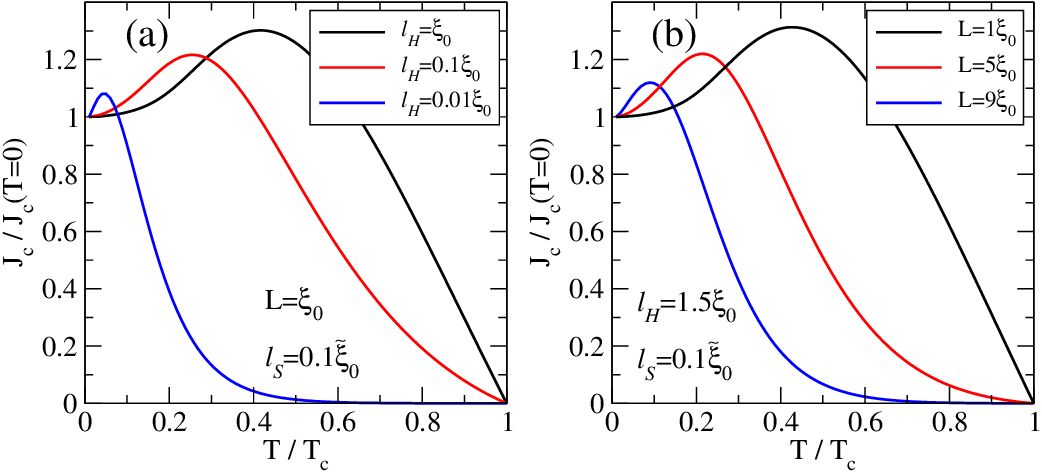}
\caption{{\bf Non-monotonic temperature dependence.} (a) The
  critical Josephson current $J_c$ has a maximum at a low temperature that for
  a specific junction length (here $L=\xi_0=v_H/2\pi T_c$) depends on the
  mean free path $\ell_H$ in the half metal. 
  $J_c$ has been normalised to the zero-temperature value. In (b) we
  show the normalised current as function of junction length $L$ for a
  fixed $\ell_H=1.5 \xi_0$. 
  When the junction becomes effectively long
  compared with the diffusive limit coherence length,
  $L=\xi_0\gg\xi_d(T)$, the current is dramatically suppressed [see
  Fig.~\ref{fig:mfp}(a)] and the peak is shifted to a lower temperature.
}
  \label{fig:Ic}
\end{figure}
In Fig. \ref{fig:mfp}(c) we show for several mean free paths the
dependence of the critical current on the junction length $L$. A rapid
exponential suppression of the effect with junction length is observed
in the diffusive limit, whereas in the moderately disordered region a
considerable effect is expected for junction lengths up to 5-10
coherence lengths.

In Fig.~\ref{fig:Ic} we discuss the influence of disorder on the temperature
dependence of the critical current. We have
normalised all $J_c(T)$ curves to their zero-temperature value.  There
is a characteristic peak appearing at a temperature below $\sim T_c/2$
as predicted for ballistic systems in Ref.~\cite{esc03}. The origin of
the peak is the factor $|\Delta |^2\vepsilon_n^2/\Omega_n^4$ in
equation~(\ref{Jc}), that results from the odd-frequency pairing amplitudes
on the {\it superconducting sides} of the interfaces being the sources
of the equal-spin correlations in the half metal, as revealed by
the factor $|\Delta | |\vepsilon_n|/\Omega_n^2$ in equation~(\ref{fhm}). 
These 
amplitudes have a dynamical degree of freedom,
that makes them less effective compared to even frequency amplitudes
when the lowest Matsubara energy $\pi T$ drops below 
$\Delta (T)$. Whereas this gives $T_{peak}\sim \Delta (T_{peak})/\pi $
for ballistic short ($L\le \xi_0$) junctions,
in the case of long or disordered junctions $T_{peak}$ is determined
by the smaller Thouless energy.
This shift to lower temperatures with decreasing $\ell_H$ is
seen in Fig.~\ref{fig:Ic}(a).
Conversely, for a particular mean free path, the peak is shifted to
lower temperatures for increasing $L$, as shown in
Fig.~\ref{fig:Ic}(b) for $\ell_H=1.5\xi_0$. For long junctions the
critical current has a characteristic exponential temperature
dependence above the peak.

\begin{figure}[t]
\includegraphics[width=1.00\columnwidth]{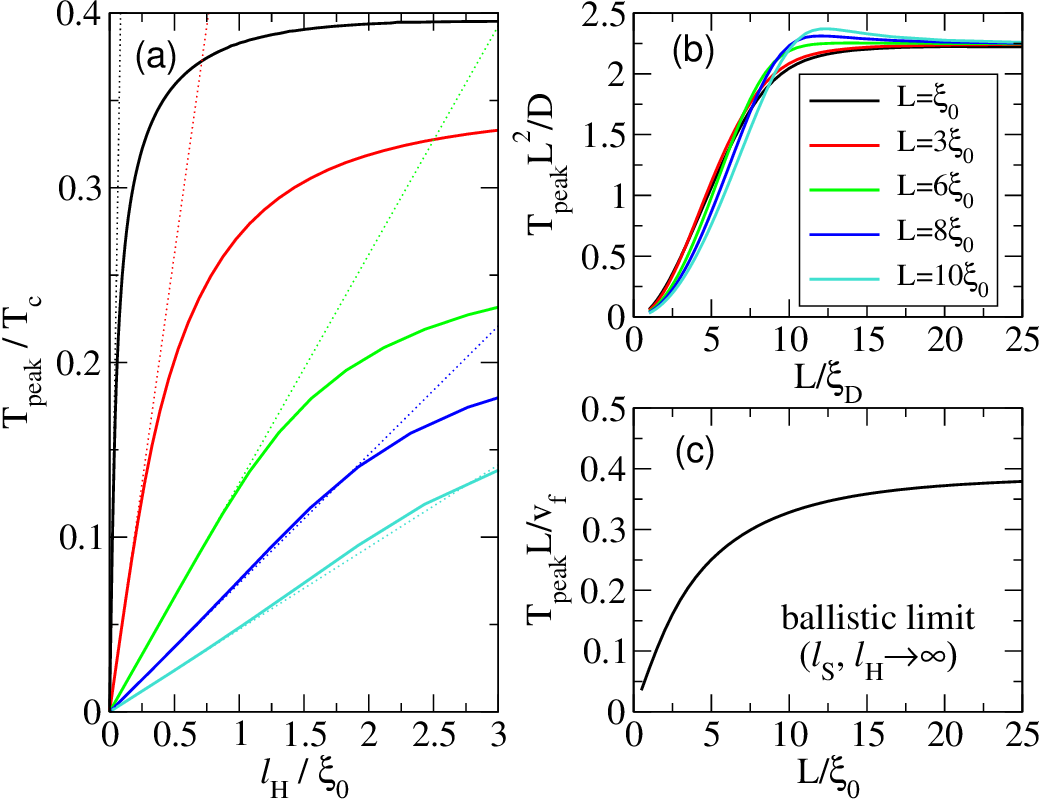}
\caption{ {\bf Peak position $T_{peak}$ in $J_c(T)$.}  (a) $T_{peak}$
  as function of the mean free path $\ell_H$ in the half metal (full lines). 
$T_{peak}$ approaches linearly zero temperature with decreasing $\ell_H$. 
The corresponding slope characterises $T_{peak}$ in the diffusive limit, and
is well described by the formula
$T_{peak}=2.25 E_{Th}$ (shown as dotted lines),
with the Thouless energy $E_{Th}=D/L^2$ and
the diffusion constant $D=v_H\ell_H/3$.
(b) Scaling plot with $T_{peak}/E_{Th}$ as a function of $L/\xi_D$ with 
$\xi_D=\sqrt{D/2\pi T_c} $.
  (c) $T_{peak}$ for the clean limit normalised to $v_H/L$ 
  as a function of $L/\xi_0$,
  showing for long junctions a scaling of $T_{peak}$ with $v_H/L$.}
  \label{fig:diffusive}
\end{figure}
In Fig.~\ref{fig:diffusive}(a) we show the peak temperature,
$T_{peak}$, for a number of junction lengths and a wide range of mean
free paths. For large mean free paths the peak position levels off
($T_{peak}$ is independent of the mean free path for clean systems),
while for small mean free paths, there is a linear relation
$T_{peak}\propto\ell_H$, implying that $T_{peak}$ vanishes in the
limit $\ell_H\to 0$. However, the diffusive limit is defined as
$\ell_H \to 0$ and $\xi_0 \to \infty $ keeping $\xi_D= \sqrt{\ell_H
  \xi_0/3}$ finite. In Fig.~\ref{fig:diffusive} (b) we show the same
data as in (a), but plotted as function of the only dimensionless
diffusive
length scale parameter $L/\xi_D$, and the position of the peak in
units of the Thouless energy $E_{Th}=D/L^2$ (diffusion constant
$D=v_H\ell_H/3$). The diffusive limit here corresponds to the curves
for which $\ell_H/\xi_0 \ll 1$, i.e. $L/\xi_D \gg L/\xi_0$. In the
limit of small diffusion length, $\xi_D \to 0$, all curves approach
the value $T_{peak}\approx 2.25 E_{Th}$. This is the region where the
diffusive (thermal) coherence length $\xi_d=\xi_D\sqrt{T_c/T}$ is
smaller than $L$. The spread of the curves in
Fig.~\ref{fig:diffusive}(b) is due to deviations from diffusive
behaviour, which leads to additional dependences on the ballistic
parameter $L/\xi_0$. For long junctions in the clean limit ($L\gg
\xi_c$), the ballistic Thouless energy $v_H/L$ is the relevant energy
scale for $T_{peak}$, as seen in Fig.~\ref{fig:diffusive}(c).

\subsection{IMPLICATIONS FOR EXPERIMENT}
Since the peak survives the crossover from the ballistic to the
diffusive limit, it serves as a true fingerprint of equal-spin triplet
correlations in the half metal. For moderately disordered junctions,
the peak is readily measurable experimentally when $L$ is of the order
of the coherence length at $T_c$. The peak is best observed in the limit
of large Thouless energy, because then the dominating energy scale is set
by the energy of the odd-frequency pairing amplitudes.
Note, finally, that in contrast to Josephson
junctions that involve weak ferromagnets, no oscillating amplitudes are
involved in half metals, thus ruling out precursor effects of a
$0$ to $\pi $ transition.

Turning to Ref.~\cite{kei06}, the low-temperature resistivity of the
CrO$_2$ material of 8.9~$\mu \Omega $~cm implies a mean free path of
$\ell_H\approx 40$~nm (note that the film thickness of 100 nm was
considerably larger than that). From band structure calculations, the
Fermi velocity for CrO$_2$ is $v_H\approx 2.2\times 10^5 $~m/s
\cite{lew97}, implying $\xi_0\approx 27 $~nm, or $\xi_0/\ell_H\approx
2/3 $ (for $T=0.3T_c$ this gives $\xi_c\approx 90 $~nm and $\xi_d
\approx 35 $~nm). The CrO$_2$ of Ref.~\cite{kei06} is therefore
moderately disordered, being in the crossover region in
Fig.~\ref{fig:mfp}(b), rather than in the clean or diffusive
limits. With the typical length of CrO$_2$ in \cite{kei06} of
$L\approx 300$~nm~$\approx 11\xi_0$, we can from Figs.~\ref{fig:Ic}(b)
and \ref{fig:diffusive}(a)
predict that if measurements are extended to lower temperatures, or a
shorter junction length is used, a peak in $J_c (T)$ should be
observed.





\section{Supplementary Material}

\renewcommand{\theequation}{\mbox{S}\arabic{equation}}
\renewcommand{\thefigure}{\mbox{S}\arabic{figure}}
\renewcommand{\thetable}{\mbox{S}\arabic{table}}

\subsection{Characterisation of interfaces with CrO$_2$}
\vspace{-0.3cm}

Although our theory is general and by no means restricted to a
specific material, it is useful to discuss several possible reasons
for the existence of misaligned moments in the interface region
relevant for experiment \cite{keiz06}.  It has been shown recently
\cite{goe07} that the CrO$_2$ films used in experiment \cite{keiz06}
have a homogeneous magnetisation profile throughout the film, but with
a magnetisation direction that is non-collinear with the magnetisation
direction of bulk CrO$_2$, as a result of the epitaxial coherence
strain at the interface between the substrate and CrO$_2$. There are
two possible ground states, and consequently the CrO$_2$ film shows a
biaxial asymmetry.  Using such biaxial CrO$_2$ films, Keizer {\it et
  al.}  reported the above mentioned long-range Josephson effect
\cite{keiz06}. The magneto-crystalline anisotropy due to spin orbit
coupling is uniaxial.

The interface between CrO$_2$ and NbTiN should be considered rather
rough, in particular for the preparation technique used in
\cite{keiz06}.  In experiment \cite{keiz06}, a well known
antiferromagnetic Cr$_2$O$_3$ layer at the CrO$_2$-surface has been
removed before contacting with the superconductor NbTiN. Here, several
possible sources for moment misalignment at the interface between the
half metallic ferromagnet CrO$_2$ and the superconductor NbTiN need to
be considered.
(A) Interface magnetic anisotropy can dominate the bulk anisotropy
leading to interface magnetism with a moment perpendicular to the bulk
magnetisation. However, in CrO$_2$ the shape anisotropy is very strong
and such interface magnetism is only plausible if disorder or
roughness of the interface weakens the exchange coupling between the
interface spins and the bulk spins.
(B) An interface with finite roughness will have a reduced exchange
coupling of interface spins to bulk spins in certain regions. As the
direction of the bulk spins is determined non-locally by an interplay
between the strain of the substrate and the uniaxial anisotropy, it
will in general deviate from the direction of the spins that do not
feel such an influence of the substrate.  As a result misalignment of
a considerable number of interface spins is probable.
(C) If clusters of spins are formed in the interface region, it has
been shown \cite{reddy00} that the magnetic coupling between the Cr
sites oscillates from antiferromagnetic to ferromagnetic and back when
successively adding an oxygen atom. Thus, the interplay between shape
anisotropy of the clusters, the crystalline anisotropy within a
cluster, and the magnetic coupling between the clusters and the bulk
magnetisation will ultimately determine the effective magnetisation of
the interface. Given the fact, that the bulk magnetisation and the
magneto-crystalline anisotropy are non-collinear, the averaged
magnetisation of the interface is expected to differ in such a case
from the bulk one, thus explicitly breaking spin rotation invariance
around the bulk magnetisation axis.
(D) Finally, for mesoscopic interfaces there is a possibility for
sample specific mesoscopic fluctuations in the spin configuration.

\subsection{Scattering matrix}
\vspace{-0.3cm}

It is known that any scattering matrix has a singular value
decomposition
\begin{equation}
\hat {\bf S}=
\left(
\begin{array}{cc} U  & 0\\ 0& \tilde U \end{array}
\right)
\left( \begin{array}{cc}
R& T
\\ T^\dagger &
-\underline{R}
\end{array} \right)
\left(
\begin{array}{cc} V  & 0\\ 0& \tilde V \end{array}
\right)
\end{equation}
where $U$ and $V$ are unitary $m\times m$ matrices, $\tilde U$ and
$\tilde V$ are unitary $n\times n$ matrices, the matrix $T$ has only
non-zero elements on the diagonal, that describe the transmission
eigenvalues (real, $0<t<1$) between the $m$ channels on one side of
the interface and the $n$ channels on the other side (the number of
transmission channels is the smaller of the two numbers $m$ and $n$).
$R=\sqrt{1-TT^\dagger }$ and $\underline R= \sqrt{1-T^\dagger T}$ are
diagonal matrices with the reflection eigenvalues. In our case, the
scattering matrix (for each fixed momentum component parallel to the
interface) connects two spin channels in the superconductor with one
spin channel in the half metal. Consequently, in our case $m=2$ and
$n=1$, and $\hat {\bf S}$ is a 3x3 matrix in spin space with
\begin{equation}
R= \left( \begin{array}{cc}
\sqrt{1-t^2}& 0 \\
0& 1  \\
\end{array} \right), \quad
\underline R =\sqrt{1-t^2}, \quad
T= \left( \begin{array}{c} 
t\\ 0 \end{array} \right).
\end{equation}
The 2x2 matrices $U$ and $V$ can be written in the form $U=
e^{i(\psi_u +\frac{\vartheta_u}{2} \vec{m}\cdot\vec{\sigma })}$,
$V=e^{i(\psi_v +\frac{\vartheta_v}{2} \vec{m}\cdot\vec{\sigma })}$, and the
scalars $\tilde U$ and $\tilde V$ are just phase factors $\tilde U=
e^{i \psi_{\tilde{u}} }$, $\tilde V= e^{i \psi_{\tilde{v}} }$. The
scalar phases $\psi_{u,v} $ and $\psi_{\tilde{u},\tilde{v}} $ are
irrelevant for the Josephson current, as they drop out of the final
expressions; thus, for convenience we put them to zero.

The unit vector $\vec{m}$ characterises the direction of the interface
magnetisation (averaged over an area comparable with the size of a
Cooper pair); the direction of the magnetisation $\bf M$ of the half
metal is the $z$-direction. In a system with spin rotation invariance
around $\bf M$ the directions $\vec{m}$ and $\vec{M}$ coincide. This
is, however, not the case if spin rotation symmetry around $\bf M$ is
broken. We denote the angle between $\vec{m}$ and $\vec{M}$ by
$\alpha$. Without loss of generality the spin quantisation axis in the
superconductor can be chosen as the $\vec{m}$-axis, and in the half
metal as the $z$-axis.  The corresponding spin rotation matrix is
$U_m=e^{-i \frac{\alpha}{2}\vec{e}_\perp\cdot\vec{\sigma}}$ with
$\vec{e}_\perp = (\vec{m} \times \vec{M})/(M \sin \alpha )$. In this
representation $U_mUU_m^\dagger =e^{i\frac{\vartheta_u}{2}\sigma_z}$
and $U_mVU_m^\dagger=e^{i\frac{\vartheta_v}{2}\sigma_z}$ become
spin-diagonal, however the transmission vector $T$ acquires a non-zero
spin-flip component, $U_mT= (t \cos \frac{\alpha }{2} , -t\sin
\frac{\alpha }{2} e^{i \pphi} )^{tr}$ where $\pphi $ is the angle of
$\vec{m}$ in the plane perpendicular to $\vec{M}$. This leads to
\begin{equation}
\label{scatt1}
\hat {\bf S}=
\left( \begin{array}{cc|c}
r_{\uparrow\uparrow}e^{\frac{i}{2}\vartheta} &
r_{\uparrow\downarrow}e^{i(\vartheta_{\uparrow\uparrow}-\vartheta_{\downarrow\uparrow})}&
t_{\uparrow\uparrow}e^{i(\vartheta_{\uparrow\uparrow}+\frac{\vartheta }{4})}\\
r_{\downarrow\uparrow}e^{-i(\vartheta_{\uparrow\uparrow}-\vartheta_{\downarrow\uparrow})}&
r_{\downarrow\downarrow}e^{-\frac{i}{2}\vartheta} &
t_{\downarrow\uparrow}e^{i(\vartheta_{\downarrow\uparrow}-\frac{\vartheta }{4})}\\
\hline
t_{\uparrow\uparrow}e^{-i(\vartheta_{\uparrow\uparrow}-\frac{\vartheta }{4})}&
t_{\downarrow\uparrow}e^{-i(\vartheta_{\downarrow\uparrow}+\frac{\vartheta }{4})}
& - r
\end{array}
\right),
\end{equation}
with $t_{\uparrow \uparrow } =
t \cos \frac{\alpha }{2}$, $t_{\downarrow \uparrow } = t\sin
\frac{\alpha }{2} $, 
$r_{\uparrow \uparrow }=\sin^2 \!\frac{\alpha }{2} +r \cos^2 \frac{\alpha }{2}$,
$r_{\downarrow \downarrow }=\cos^2 \frac{\alpha }{2} +r \sin^2 \!\frac{\alpha }{2}$,
$r_{\uparrow \downarrow }=r_{\downarrow \uparrow }=-\frac{1-r}{2}\sin \alpha $,
$r=\sqrt{1-t^2}$,
$\vartheta =\vartheta_u+\vartheta_v$,
$\vartheta_{\uparrow \uparrow} =\frac{\vartheta_u-\vartheta_v}{4}$,
and $\vartheta_{\downarrow \uparrow }= \pi+\pphi+\vartheta_{\uparrow
  \uparrow}$,
which reduces for $t\ll 1$ 
to the scattering matrix given in equation~(1) of the paper. 

\subsection{Methods}

We obtain the Josephson current as function of impurity concentration,
temperature and junction length using the quasiclassical Green's
functions technique \cite{Eilenberger,Larkin}.  The Green's functions
$\hat{g}(\vec{p}_F,{\bf R},\vepsilon_n)$ depend on the spatial
coordinate ${\bf R}$, Matsubara energy $\vepsilon_n=(2n+1)\pi T$, and
the momentum direction on the Fermi surface $\vec{p}_F$.  In the
superconductors, the propagator $\hat g$ is a 4x4 matrix in combined
spin and particle-hole space,
\begin{equation}\label{eq:greenS}
\hat g^S = \left(
\begin{array}{cc}
g_s + \vec g_t\cdot\vec\sigma & (f_s + \vec f_t\cdot\vec\sigma)i\sigma_y \\
( \tilde f_s + \tilde{\vec f}_t \cdot\vec\sigma^\ast )i\sigma_y  &
\tilde g_s + \tilde{\vec g}_t \cdot\vec\sigma^\ast
\\
\end{array}
\right),
\end{equation}
where $f_s$ and $\vec f_t$ are singlet and triplet pairing amplitudes,
$g_s$ and $\vec g_t$ are spin scalar and spin vector parts of the
diagonal Green function, and the vector
$\vsigma=(\sigma_x,\sigma_y,\sigma_z)$ is composed of Pauli spin
matrices. The hole amplitudes are related to the particle amplitudes
by the symmetry 
$\tilde f(\vec{p}_F,\vepsilon_n)=f(-\vec{p}_F,\vepsilon_n)^\ast$.
In the half metal, only conduction electrons with spin up exist, and
the propagator is a 2x2 matrix in particle-hole space,
\begin{equation}\label{eq:greenF}
\hat g^{H} = \left(
\begin{array}{cc}
g_{\uparrow\uparrow} & f_{\uparrow\uparrow} \\
\tilde f_{\uparrow\uparrow} & \tilde g_{\uparrow\uparrow} \\
\end{array}
\right).
\end{equation}
The propagators are connected at the interfaces via the scattering
matrices given in equation~(1) of the paper.

The transport equation governing the supercurrent in the heterostructure
is given by the Eilenberger equation for the propagator $\hat g$.
Impurities are treated in the Born approximation
using a life time $\tau_S$ of quasiparticles in the superconductor, and
a life time $\tau_H$ in the half metal. The corresponding mean free paths
are $\ell_S=v_S\tau_S$, $\ell_H=v_H\tau_H$, with the Fermi velocities
$v_S$ and $v_H$ in the two materials.
The equation of motion for the 4x4 Green's function 
in the superconductors reads,
\begin{equation}
\label{qcl1}
iv_{S} \, \mu \, \partial_x \hat g + \left[
i\vepsilon_n \hat \tau_3 -\hat \Delta - \frac{1}{2\pi\tau_S } \langle \hat g \rangle , \hat g
\right]=\hat 0,
\end{equation}
where $\mu=\cos (\theta_p )$,
$\theta_p$ is the angle between the Fermi velocity and the $x$-axis,
$\hat\tau_3$ is the third Pauli matrix in particle-hole space, and 
$\hat \Delta = \Delta \hat 1 i\sigma_y$
is the singlet order parameter.
The average $\langle \cdots \rangle = \int 
\frac{d \cos (\theta_p)d\pphi_p}{4\pi} \cdots $ 
is over all momentum directions.
There is an analogous equation for the
2x2 Green's function in the half metal,
\begin{equation}
\label{qcl2}
iv_{H} \, \mu \, \partial_x \hat g + \left[
i\vepsilon_n \hat \tau_3  - \frac{1}{2\pi\tau_H } \langle \hat g \rangle , \hat g
\right]=\hat 0.
\end{equation}
Equations (\ref{qcl1})-(\ref{qcl2}) are supplemented with the
normalisation condition $\hat g^2 =-\pi^2 \hat 1$.

\begin{table}[t]
\begin{tabular}{|r|c|c|}
\hline
\hline
symmetry & even frequency & odd frequency \\
\hline
even parity&
$ f_s^{s}=
\frac{\pi \Delta \Omega_n \left(1-\sin^2 \!\frac{\vartheta}{2}\right)}{\Omega_n^2-|\Delta |^2 \sin^2 \!\frac{\vartheta}{2} } $ & 
$ f_{t_0}^s= 
\frac{ i\pi \Delta  \vepsilon_n \; \frac{1}{2}\sin \vartheta }{\Omega_n^2-|\Delta |^2 \sin^2 \!\frac{\vartheta}{2}} $ \\
\hline
odd parity &
$f_{t_0}^a= 
\frac{-i\pi \Delta 
\Omega_n s_\mu \; \frac{1}{2}\sin \vartheta }{\Omega_n^2-|\Delta |^2 \sin^2 \!\frac{\vartheta}{2}} $ &
$f_s^{a}= 
\frac{\pi \Delta \vepsilon_n s_\mu \sin^2 \!\frac{\vartheta}{2}
}{\Omega_n^2-|\Delta |^2 \sin^2 \!\frac{\vartheta}{2}} $\\
\hline
\hline
\end{tabular}
\caption{(Supplementary Table) Symmetry components of the interface amplitudes 
in the superconductors for the the clean limit, assuming a constant 
singlet order parameter $\Delta$ and small tunnelling amplitudes ($t\ll
1$). Here, $\Omega_n=\sqrt{|\Delta|^2+\vepsilon_n^2}$ and 
$s_\mu={\rm sgn} (\mu )$.}
\label{supplTable0}
\end{table}

We linearise the above equations for
small triplet components in the superconductor ($f_{t_0}$) and in the
half metal ($f_{\uparrow\uparrow}$). 
From the clean limit solutions
for the interface amplitudes in the superconductors for $t\ll 1$
and for a constant singlet order parameter $\Delta$
(shown in Supplementary Table~\ref{supplTable0}), we see that
it is necessary, in order to ensure small $f_{t_0}$, to
neglect higher order terms in the parameter $\sin (\vartheta )\approx\vartheta $.
We will do so in the following.
The correction to $f_s^s$ are then of order
$\sin^2 \!\frac{\vartheta }{2}$, leaving the order parameter unaffected
in leading order in $\vartheta $.
The normalisation condition is
used to eliminate the diagonal part of $\hat g$ in favour of a
coupled set of equations for $f$-functions with positive and negative
momentum directions. We decouple the system of differential equations
by introducing the new triplet functions $\ppsi_{0j} $ and $\ppsi_{j}$
in equations~(2)-(3) in the paper.

The solutions for the functions $\ppsi^{s,a}_{0j}$, appearing in the
ansatz Eq.~(2) for 
the superconductors, are given by
\begin{eqnarray}
\label{iesc1}
\ppsi_{01}^{s} (x) &= & \frac{s_\vepsilon }{2} B_{01 }(x)  +
\int_{-\infty}^{x_1} dx' \; 
\frac{K_{1}(x,x')}{2|\mu |\ell_S}  \langle \ppsi^s_{01}(x')\rangle  
\\
\label{iesc2}
\ppsi_{02}^{s} (x) &= & \frac{s_\vepsilon }{2} B_{02}(x)  +
\int_{x_2}^{\infty } dx' \;\frac{K_{2}(x,x')}{2|\mu |\ell_S}  \langle \ppsi^s_{02}(x')\rangle  
\qquad
\end{eqnarray}
with $s_\vepsilon=\mbox{sgn}(\vepsilon_n )$,
$ B_{0j }(x)=e^{-|x-x_j |/\xi_S |\mu | } $,
$ K_{j }(x,x')= e^{-|x-x'|/\xi_S |\mu |}+
e^{-(|x'-x_j |+|x-x_j |)/\xi_S |\mu |)} $, and
$\xi_S = v_{S}/(2\Omega_n+\tau_S^{-1})$ with $\Omega_n=\sqrt{\vepsilon_n^2+|\Delta|^2}$.
The momentum-antisymmetric parts are obtained by using the identity
$\ppsi_{0j}^{a}=-\mu s_\vepsilon \xi_S \partial_x \ppsi_{0j}^{s} $.

After using the boundary conditions with the scattering matrix 
$\hat {\bf S}$ for $\hat g^S $ and $\hat g^H$, we obtain
the solutions 
$\ppsi^{s,a}_{j}$
in the half metal,
\begin{eqnarray}
\label{iehm}
\ppsi^{s}_{j} (x) &= &
\frac{1}{1-e^{-2L/\xi_{H} |\mu |}}\Big( \frac{s_\vepsilon }{2} B_{j }(x) +
\Big. \nonumber \\
&&\Big. 
\int_{x_1}^{x_2} dx' \;
\frac{K(x,x')}{2|\mu |\ell_H } \langle 
\ppsi^s_{j}(x')
\rangle \Big)
\end{eqnarray}
with $\xi_{H}=v_{H}/(2|\vepsilon_n|+\tau_H^{-1} )$, 
$L=x_2-x_1$,
$ s_\vepsilon B_1(x)= \ppsi^s_{01}(x_1) \left(e^{-(x-x_1 )/\xi_{H} |\mu |}+ 
e^{-(L+x_2-x)/\xi_{H} |\mu |}\right) $, $
s_\vepsilon B_2(x)= \ppsi^s_{02}(x_2) \left(e^{-(x_2-x )/\xi_{H} |\mu |}+ 
e^{-(L+x-x_1)/\xi_{H} |\mu |}\right) $, $
K(x,x')= e^{-|x-x'|/\xi_{H} |\mu |}+ e^{-(2L-|x-x'|)/\xi_{H} |\mu |}+ 
e^{-(x+x'-2x_1)/\xi_{H} |\mu |}+ e^{-(2x_2-x-x')/\xi_{H} |\mu |} $.
The momentum-antisymmetric parts are obtained by using the identity
$\ppsi_{j}^{a}=-\mu s_\vepsilon \xi_{H} \partial_x \ppsi_{j}^{s} $.

The integral equations (\ref{iesc1})-(\ref{iehm}) 
for $\langle \ppsi^s (x) \rangle $
are solved by replacing $\ppsi $ on a spatial grid
with a piecewise linear function. 
Exact integration of the resulting expressions
reduces the problem to a simple matrix inversion.
The angular averages
can be performed analytically and lead to exponential integrals.
This procedure is necessary because the integration
kernel decays on a different length scale compared with $\ppsi $ in the
diffusive limit.

The current density in the half metal
is given by the diagonal Green's function,
\begin{equation}
J_x(x)=e v_{H} N_{H }
T\sum_{\vepsilon_n} \langle \mu g_{\uparrow\uparrow}(\mu,\vepsilon_n,x) \rangle,
\end{equation}
where $N_{H}$ is the density of states at the Fermi level for the
conducting spin band in the half metal, and $e$ is the electronic charge.
It can be shown from the transport equations (\ref{qcl1}) and (\ref{qcl2}) that
the current density $J_x$ does in fact not depend on the spatial coordinate,
in agreement with the continuity equation that expresses particle conservation.

Using the normalisation condition in the half metal,
$g_{\uparrow\uparrow}^2=-\pi^2-
f_{\uparrow\uparrow } \tilde f_{\uparrow\uparrow}$, for small
triplet amplitudes, one obtains in leading order
$g_{\uparrow\uparrow}=-i\pi \mbox{sgn} (\vepsilon_n ) (1+ 
f_{\uparrow\uparrow } \tilde f_{\uparrow\uparrow}/2\pi^2)$,
leading to
\begin{equation}
J_x=\frac{ev_{H} N_{H }}{2\pi i}T\sum_{\vepsilon_n} \langle \mu 
f_{\uparrow\uparrow }(\mu,\vepsilon_n,x) \tilde f_{\uparrow\uparrow}(\mu,\vepsilon_n,x)
\rangle \mbox{sgn} (\vepsilon_n ).
\end{equation}
It is instructive to decompose the anomalous propagators into their 
symmetric and antisymmetric part $f_{\uparrow\uparrow}^{s,a}$ with respect to $\mu $. 
Doing this and using the fundamental symmetries 
$f_{\uparrow\uparrow}^s(-\vepsilon_n)=-f_{\uparrow\uparrow}^s(\vepsilon_n)$, $f_{\uparrow\uparrow}^a(-\vepsilon_n)=f_{\uparrow\uparrow}^a(\vepsilon_n)$ we 
arrive at
\begin{equation}
\label{JJ}
J_x=-\frac{2ev_{H}N_{H }}{\pi }T\sum_{\vepsilon_n>0} 
\int_0^1 d\mu \; \mu \mbox{Im} (f_{\uparrow\uparrow}^s f_{\uparrow\uparrow }^{a\ast }).
\end{equation}
Substitution of equation~(3) of the paper into equation~(\ref{JJ}) leads to
equations~(6)-(7) of the paper.

To study the symmetry properties of the Josephson current, we expand
the pairing amplitudes in Legendre polynomials. Writing
$\overline \ppsi_j(\mu ) = A_j(\mu )\ppsi_j (\mu )$, the expansion is
$\overline \ppsi_j(\mu) = \sum_{l=0}^\infty P_l(\mu ) \overline \ppsi_{j,l}$, 
with components $\overline\ppsi_{j,l}=(2l+1) \langle P_l(\mu ) \overline \ppsi(\mu )\rangle $.
Using that
\begin{eqnarray}
&&\int_0^1 d\mu \; \mu 
(\overline\ppsi^s_{2} \overline\ppsi^a_{1} - \overline\ppsi^s_{1} \overline\ppsi^a_{2} ) = \nonumber \\
&&\sum_{l=0}^\infty 
\frac{(-1)^l(l+1)}{(2l+1)(2l+3)}  \left( \overline\ppsi_{2,l} \overline\ppsi_{1,l+1}-\overline\ppsi_{1,l}\overline\ppsi_{2,l+1}
\right),~~~~
\end{eqnarray}
we can bring equation~(7) in the paper to the form $J_c=\sum_{l=0}^\infty
J_{c;l,l+1}$.

The various length scales appearing in the paper are listed in
Supplementary Table~\ref{supplTable}.
\begin{table}[t]
\begin{tabular}{ll}
\hline
\hline
Superconductor side: & $\ell_S=v_S\tau_S$\\
& $\xi_S=\frac{v_S}{2\Omega_n+\tau_S^{-1}}$\\
                     & $\tilde\xi_0=\frac{v_S}{2\pi T_c}$\\
\hline
Half-metallic side: &$\ell_H=v_H \tau_H$\\
& $\xi_H=\frac{v_H}{2|\vepsilon_n|+\tau_H^{-1}}$\\
                    & $\xi_c=\frac{v_H}{2\pi T}$\\
                    & $\xi_0=\frac{v_H}{2\pi T_c}$\\
                    & $\xi_d=\sqrt{\frac{\ell_H\xi_c}{3}} = \sqrt{\frac{D}{2\pi T}}$\\
                    & $\xi_D=\sqrt{\frac{\ell_H\xi_0}{3}}= \sqrt{\frac{D}{2\pi T_c}}$\\
\hline
\hline
\end{tabular}
\caption{(Supplementary Table) Collection of various length scales that enter in the problem.
  Here $v_S$ and $v_H$ are Fermi velocities in the superconductor and the half-metal,
  respectively, while $\tau_S$ and $\tau_H$ are the impurity scattering times
  in the two materials, and $D$ is the diffusion constant in the half metal.}
\label{supplTable}
\end{table}

\subsection{Supplementary Results}

\setcounter{figure}{0}
\begin{figure}[t]
\includegraphics[width=\columnwidth]{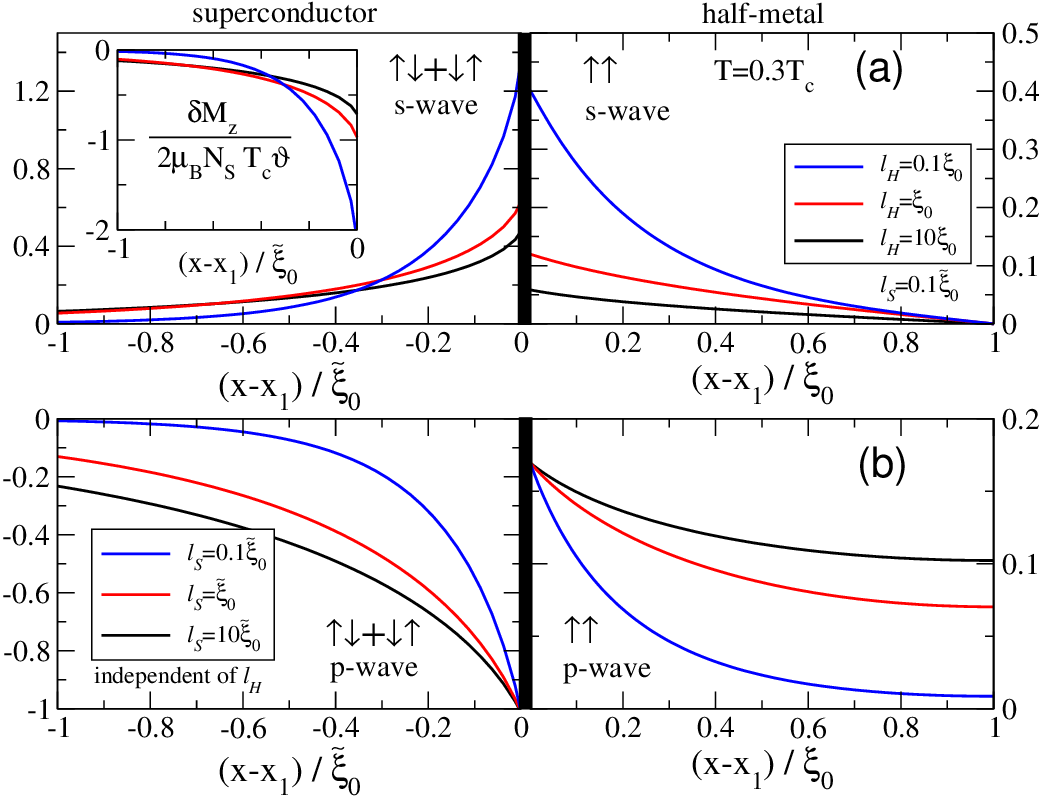}
\caption{(Supplementary Figure)
Spatial dependences of pairing amplitudes in the superconductor
(in units of $\sin (\vartheta ) |\Delta | e^{i\chi_j}$) and
in the half metal (in units of $iA_j|\Delta | e^{i\overline\chi_j}$,
length $L=2\xi_0$, only left half shown) 
for a $\pi$-junction. 
The odd-frequency $s$-wave triplet
amplitude is shown in the upper panel, the even-frequency $p$-wave triplet
amplitude in the lower panel. 
In the inset we show the induced
spin polarisation of quasiparticles near the interface in the superconductor.
}
\label{fig:Ffun}
\end{figure}
In Supplementary Fig.~\ref{fig:Ffun} we show solutions of the integral
equations in the superconductor, equations~(\ref{iesc1})-(\ref{iesc2}), and
in the half metal, equation~(\ref{iehm}), for several impurity
concentrations ranging from the ballistic limit to the diffusive
limit.  
The triplet amplitudes in the superconductor 
are $m=0$ with respect to the interface moments, and 
in the half metal are equal-spin $m=1$ amplitudes with respect
to the bulk magnetisation axis ${\bf M}$. The interface moments
can be misaligned with respect to ${\bf M}$
as result of a (spontaneous or induced)
breaking of spin-rotation symmetry around ${\bf M}$
at the interface.
The symmetry components in the half metal are defined as 
\begin{eqnarray}
F_j^{s}(x)&=& T\sum_{\vepsilon_n>0}
  \langle f_{\uparrow\uparrow j} (x)\rangle \frac{|\Delta |}{\Omega_n}, 
\end{eqnarray}
\begin{eqnarray}
F_j^{p}(x)&=& \frac{1}{3}T\sum_{\vepsilon_n>0}
    \langle \mu f_{\uparrow\uparrow j} (x)\rangle \frac{|\Delta |}{\Omega_n}.
\end{eqnarray}
In the superconductor an analogous definition holds for the 
$m=0$ components $F_{t_0j}^s$ and $F_{t_0j}^p$.
In the left panels of Supplementary Fig.~\ref{fig:Ffun}
we vary the superconducting mean free path
$\ell_S$, and in the right panels the half metal mean free path
$\ell_H$ for fixed $\ell_S$.  
Whereas the interface value of 
$F_{t_0j}^p$ in the superconductor does not change with
varying mean free path, the interface value of $F_{t_0j}^s$ increases
with decreasing mean free path in the superconductor, $\ell_S$, as
$1/\sqrt{\ell_S}$ until it reaches values comparable with the singlet
amplitude. Their decay length in the superconductors decreases, and
changes from $(\xi_S^{-1} +\ell_S^{-1})^{-1}$ in the ballistic limit
to $\sqrt{\xi_S \ell_S/3 }$ in the diffusive limit.
An analogous picture is seen on the half-metallic side.

In the inset we also show the induced
spin polarisation in the superconductors.  It is calculated from
the diagonal part of the Green's function, given (for small $\vartheta_j $) by
\begin{eqnarray}
g_{zj}(x)=-\pi  \frac{|\Delta|^2 \sin \vartheta_j }{\vepsilon_n^2+|\Delta |^2} 
\mbox{sgn}(\vepsilon_n) \ppsi_{0j}^s(x).
\end{eqnarray}
There is, consequently, a surface spin polarisation in the superconductor, that
in the clean limit is given by
\begin{equation}
\langle g_{zj}(x)\rangle =-\; \frac{\pi |\Delta|^2 }{2(\vepsilon_n^2+|\Delta |^2)}
\langle \sin \vartheta_j e^{-|x-x_j|/\xi_S |\mu |}\rangle .
\end{equation}
The induced spin magnetisation is then
\begin{equation}
\delta M_z(x)=2\mu_B N_S T\sum_{\vepsilon_n} \langle g_{zj}(x)\rangle ,
\end{equation}
where $\mu_B$ is the Bohr magneton and $N_S$ is the density of states
in the superconductor.

\begin{figure}[t]
  \includegraphics[width=\columnwidth]{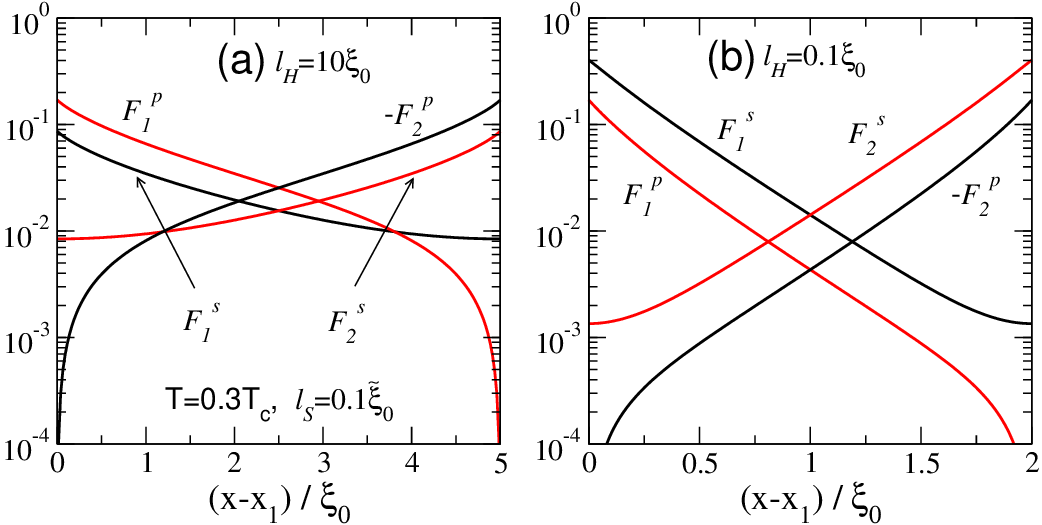}
  \caption{
(Supplementary Figure)
Role of triplet correlation functions in the half metal. 
The contribution to the Josephson current from the $s$-wave 
$\uparrow\uparrow$-triplet
correlation function always enters as a product with the
$p$-wave $\uparrow\uparrow$-triplet, 
with one of the two originating from the left superconductor
and the other from the right. For clean half metals, shown in (a),
the $p$-wave component is larger than the $s$-wave. For more dirty
structures, shown in (b), the $p$-wave component is suppressed
compared with the $s$-wave and the Josephson current is suppressed
accordingly. Amplitudes are
plotted in units of $iA_j|\Delta | e^{i\overline\chi_j}$.
}
\label{fig:f_lr}
\end{figure}
In Supplementary Fig.~\ref{fig:f_lr} we present an analysis of the
spatial dependences of the odd-frequency $s$-wave and even-frequency
$p$-wave pairing amplitudes in the half metal. 
We show for the
ballistic case ($\ell_H=10\xi_0$) and for the diffusive case
($\ell_H=\xi_0/10$) the pairing amplitudes induced from the left and
right interfaces.  By multiplying the two black curves with each other
and the two red curves with each other, and summing the two
contributions, we obtain a quantity related to the $s\cdot p$
contribution to the Josephson current [see equation~(7)].  For ballistic
systems the $p$-wave amplitudes are larger than the $s$-wave
amplitudes near the interfaces, while the opposite holds for diffusive
systems.
Amplitudes for fixed frequency give a similar picture.

\end{document}